%% file: 6.tex
\documentclass[runningheads]{llncs}
\usepackage[T1]{fontenc}
\usepackage{graphicx}
\usepackage{paralist}

\usepackage{soul, xcolor}

\usepackage{amsfonts}
\usepackage{amsmath}
\input{mathcommends}

\usepackage{booktabs}   
\usepackage{multirow}
\usepackage{fmtcount}

\begin{document}
\title{%Neovascular age-related macular degeneration activity change detection and prediction in retinal optical coherence tomography \\ OR \\
%Detecting and predicting neovascular age-related macular degeneration activity change  in retinal optical coherence tomography \\ OR \\
Automatic detection and prediction of nAMD activity change  in retinal OCT using Siamese networks and Wasserstein Distance for ordinality}
\author{Taha Emre$^{1*}$ \and Teresa Araújo$^{2*}$ \and Marzieh Oghbaie$^{2*}$
Dmitrii Lachinov\inst{2} \and
Guilherme Aresta\inst{2} \and
Hrvoje Bogunovi\'c\inst{2}}
%index{Emre, Taha}
%index{Araújo, Teresa}
%index{Oghbaie, Marzieh}
%index{Lachinov, Dmitrii}
%index{Aresta, Guilherme}
%index{Bogunotinvic, Hrvoje}
%
\authorrunning{Emre et al.}
\institute{Dept. of Ophthalmology and Optometry, Medical University of Vienna, Austria \and 
Christian Doppler Laboratory for Artificial Intelligence in Retina, Institute of Artificial Intelligence, Center for Medical Data Science, Medical University of Vienna, Austria}
\maketitle              % typeset the header of the contribution
\def\thefootnote{*}\footnotetext{These authors contributed equally to this work}\def\thefootnote{\arabic{footnote}}
\begin{abstract}
%The abstract should briefly summarize the contents of the paper in 150--250 words.

%Neovascular age-related macular degeneration (nAMD) is a leading cause of vision loss among older adults, where disease activity detection and progression prediction are critical for nAMD management in terms of timely drug administration and improving patient outcomes. Recent advancements in deep learning offer a promising solution for predicting the change in AMD from optical coherence tomography (OCT) volume of the retina. In this work, we proposed deep learning models for the two tasks of the public MARIO Challenge at MICCAI 2024, designed to detect and forecast changes in nAMD severity with longitudinal retinal OCT. For the first task, we employ a Vision Transformer (ViT) based Siamese Network to detect changes in AMD severity by comparing scan embeddings of a patient from different time points. To train a model to forecast the change after 3 months, we exploit, for the first time, an Earth Mover Distance-based loss to harness the ordinal relation within the severity change classes. Both models ranked high on the preliminary leaderboard, demonstrating that their predictive capabilities could facilitate nAMD treatment management.%timely treatment planning while reducing the burden on hospitals. 
Neovascular age-related macular degeneration (nAMD) is a leading cause of vision loss among older adults, where disease activity detection and progression prediction are critical for nAMD management in terms of timely drug administration and improving patient outcomes. Recent advancements in deep learning offer a promising solution for predicting changes in AMD from optical coherence tomography (OCT) retinal volumes. In this work, we proposed deep learning models for the two tasks of the public MARIO Challenge at MICCAI 2024, designed to detect and forecast changes in nAMD severity with longitudinal retinal OCT. For the first task, we employ a Vision Transformer (ViT) based Siamese Network to detect changes in AMD severity by comparing scan embeddings of a patient from different time points. To train a model to forecast the change after 3 months, we exploit, for the first time, an Earth Mover (Wasserstein) Distance-based loss to harness the ordinal relation within the severity change classes. Both models ranked high on the preliminary leaderboard, demonstrating that their predictive capabilities could facilitate nAMD treatment management.$^{**}$%timely treatment planning while reducing the burden on hospitals. 
\def\thefootnote{**}\footnotetext{https://github.com/EmreTaha/Siamese-EMD-for-AMD-Change}\def\thefootnote{\arabic{footnote}}
\keywords{Longitudinal change detection \and Age-related macular edema \and Siamese networks \and Ordinal classification.}
\end{abstract}

\section{Introduction}

% ----- AMD progression 
Neovascular age-related macular degeneration (nAMD) is a progressive exudative disease characterized by the accumulation of fluid in the macula, which can significantly impair vision function~\cite{guymer2023age}. Anti-VEGF treatments have shown great efficacy in mitigating AMD progression, and the positive effects are optimized by reducing the time from fluid onset to treatment, and thus regular follow-up is crucial for successful patient outcomes. 
The presence and extent of exudative signs, such as intraretinal and subretinal fluid, which are best visible on optical coherence tomography (OCT) images, are relevant markers for anti-VEFG  administration~\cite{schmidt2016paradigm}. Thus, predicting and accurately detecting changes in neovascularization activity are pivotal tasks for treatment management.
% - summary:
%\hl{Summary of last par. if needed:}
%Age-related macular degeneration (AMD) is a progressive condition that leads to macular deterioration and vision loss \hl{citation?}. Anti-VEGF treatments effectively slow AMD progression, especially when started early and followed up regularly \hl{citation?}. Key indicators for treatment, exudative signs such as subretinal and intraretinal fluid, are detectable through optical coherence tomography (OCT) imaging, making accurate monitoring essential for treatment decisions \hl{citation?}.  
% 
%\HBc{Start talking about MARIO challenge here which was designed to address these tasks with AI support.}
%\HBc{I meant to move the 1.4 part here but it can also stay as is.}
The "Monitoring Age-related Macular Degeneration Progression In Optical Coherence Tomography" (MARIO) Challenge, organized as part of MICCAI 2024~\cite{mario2024}, aims the development of automatic methods for nAMD progression assessment and evolution prediction. 

\subsection{Longitudinal change detection in retinal OCT}

%  ------ Motivation for change detection in OCT, particularly in nAMD ?
%Using medical images to evaluate diseases and assess changes over time is a routinal and important task in clinical decision making.
%Understanding how a patient's condition has evolved is crucial to guide the patient's clinical management.

% ----- Motivation for use of automatic methods DL 
%Manually assessing the changes between image pairs can be a very time consuming and cumbersome task. As such, the development of automated systems to detect meaningful changes between images play an important role in reducing the work load of specialists. 
%Despite the great progress in machine learning solutions, and in particular deep learning (DL)-based, to automate OCT diseases diagnostic tasks, little attention has been given to the change between sequential OCTs. 
% - summary: 
The manual assessment of changes between image pairs is time-consuming and challenging, highlighting the need for automated systems to detect meaningful changes and reduce specialists' workload. Although machine learning, particularly deep learning, has advanced in automating OCT disease diagnostics, there has been little focus on detecting changes between sequential OCT images.

% ----- Current DL approaches for change detection in medical imaging

In medical image analysis, current deep learning methods for change detection usually fall into two categories:
\begin{inparaenum}[1)]
\item Siamese networks~\cite{Li2020,Li2020covid,To2021}, or
\item Graph-based methods~\cite{Karwande2022,Rochman2023}
\end{inparaenum}.
%Li2020 - siamese CL ------------------
%Particularly, Li \textit{et al.}~\cite{Li2020} formulate the change detection task as a distance metric-learning problem, proposing a Siamese neural network to evaluate change between longitudinal patient visits on a continuous spectrum.
%A contrastive loss function is applied between features of two images, with ground truth label \textit{change} or \textit{no change}. The network, thus, outputs a pairwise distance between patient images from 2 time points. 
%Occlusion sensitivity is applied for assessing the location of the relavant image patches for the change prediction. 
%The authors tested on retinal fundus images (diabetic retinopathy) and knee x-rays (osteoarthritis). Similarly, in a subsequent work, Li \textit{et al.}~\cite{Li2020covid} applied a Siamese NN-based approach for COVID-19 pulmonary disease severity tracking on chest x-rays. 
% - summary: 
In Li \textit{et al.}~\cite{Li2020}, change detection is approached as a metric-learning problem, using a Siamese neural network to assess changes between two patient visits. A contrastive loss function is applied between features of two images, labeled as \textit{change} or \textit{no change}, to output a pairwise distance between images from two time points. This method was tested on retinal photographs (diabetic retinopathy of prematurity) and knee x-rays (osteoarthritis).  In Li \textit{et al.}~\cite{Li2020covid}, the same authors used a similar approach to track COVID-19 pulmonary disease severity in chest x-rays.

%- xray work with graphs -----------------
%Distinctively, Karward \textit{et al.}~\cite{Karwande2022} propose  a graph-based anatomy-aware model for tracking longitudinal relations between chest x-rays. It utilizes both local and global anatomical information to output localized comparisons between two sequential x-rays, outperforming Siamese-based baselines. However, this type of approach is more suitable to cases where the structures of interest are contained in a region that is not sigificantly deformable, and thus of well defined location (e.g., chest x-rays or brain CTs), and not that suitable for cases such as retinal OCT, where pathologies can cause significant deformations in the tissues.
% - summary: 
In contrast, Karward \textit{et al.}~\cite{Karwande2022} proposed a graph-based, anatomy-aware model for tracking changes between chest X-rays by using both local and global anatomical information. This model provides localized comparisons between sequential X-ray images and outperformed Siamese-based models. However, this type of approach is more suitable when the structures are within a rigid region, and thus have a stable anatomical location (e.g., chest x-rays), and are not that suitable for retinal OCT, where pathologies can cause significant tissue deformations.
%- summary of last par: 
%However, it is most effective for images where the structures are relatively stable and well-defined, such as chest x-rays or brain CTs, and less suitable for images like retinal OCT, where significant tissue deformations can occur.

%"as estruturas de interesse estão contidas numa região não significativamente deformável, e por isso de localização bem definida"

% works that require pixel wise annotations
%There are also approaches that require pixel-wise annotations of lesions and/or organs in the images, such as \cite{Szeskin2023}, which are not always available and require additional annotation effort.\hl{change this, not very good}

%\hl{see if mention fluidregnet or not - maybe not, too recent}

% ====================================================================
\subsection{Prediction of nAMD evolution within 3 months on OCT}
nAMD severity change prediction is crucial for patient follow-up scheduling and timely drug (anti-VEGF) administration. In this aspect, it is essential to have a predictive model that can assess the change in AMD disease activity within a meaningful time-frame from an OCT scan acquired at a visit. Prior work has largely focused on a related task of intermediate AMD progression prediction%, where the time-frame is generally set to a clinically-meaningful interval of 6 to 12 months
~\cite{emre2022tinc,chakravarty2024morph,rivail2019modeling,yim2020predicting}. The main challenge of training a predictive model of disease progression and treatment response is the training data. Since, the nature of the task is temporal, it is crucial to have follow-up visits from a patient to create a longitudinal dataset. In~\cite{emre2022tinc,chakravarty2024morph,rivail2019modeling}, they used longitudinal self-supervised learning methods to learn the temporal relations. Additionally, most of the available OCTs have no observable change within short time windows, resulting in severe class imbalance, requiring specialized deep learning loss terms to address this~\cite{lin2017focal}.

% ----- Contributions ---------------------------------------------------
\subsection{Contributions}
We contribute to the state of the art in longitudinal retinal OCT assessment in two ways. First, we propose a Siamese-based approach, SiamRETFound, relying on an OCT foundation model (RETFound~\cite{Zhou2023}) and additional pretraining for learning to estimate clinically-relevant changes between two patient B-scans. %The model is tested in the MARIO challenge data, and trained on MARIO data after pretraining on public OCT datasets.
Second, we propose a classifier model that predicts the nAMD evolution within 3 months. Specifically, our novel loss has 2 parts: a focal loss to address the severe class imbalance, and an Earth Mover's Distance-based (EMD) loss to harness the ordinal relation within the severity change classes. To our knowledge, this is the first study to use EMD for predicting disease severity evolution.

%Also references to Change detection in satellite imgs ? - SEE HOW TO INCORPORATE 

\subsection{MARIO Challenge}
%Task 1
%We assess our methods on the two tasks of the "Monitoring Age-related Macular Degeneration Progression In Optical Coherence Tomography" (MARIO) Challenge~\cite{mario2024}.
We assess our methods on the two tasks of the MARIO challenge. 
%The MARIO challenge consisted of two separate tasks.
%The main goal of MARIO task 1 was to develop algorithms to recognize the evolution of neovascular activity in OCT scans of patients with exudative AMD, for the purpose of improving the planning of anti-VEGF treatments. The system should be able to classify evolution between pairs of 2D slices/B-scans from OCT volumes of two different patient visits in one of the following categories: 
%\begin{inparaenum}[1)]
%    \item  Reduced,
%    \item Stable,
%    \item Worsened, or
%    \item Other (Uninterpretable/Appeared and eliminated). 
%\end{inparaenum}
%Furthermore, the usage of private data was not allowed. 
%
% summary:
Task 1 (T1) aims at developing algorithms to classify changes in disease activity from pairs of retinal OCT B-scans from two visits of patients with nAMD undergoing anti-VEGF treatment, to support decision-making. The classification categories were \textit{Reduced, Stable, Worsened}, or \textit{Other} (Uninterpretable). Based on our qualitative assessment of the provided training set, \textit{Other} seemed to be associated with scenarios that prevented proper clinical assessment, such as the presence of noise, obscured regions, vertical flipping of the scan, poor alignment between image pairs, etc. (Fig.~\ref{fig:exs_preds}: (e)-(g)). However, no objective criteria was provided, and there were a few nuanced cases (Fig.~\ref{fig:exs_preds}: (g)).

In Task 2 (T2), the goal is to train a predictive model that can accurately predict the change in disease activity within 3 months, given a single B-scan. The prediction categories are \textit{Reduced}, \textit{Stable} and \textit{Worsened}.%, similarly to T1. 
Even though the labels and the prediction are on B-scan level, labels are consistent within a volume.%, unlike T1.
% ====================================================
% ====================================================

\section{Method} % (Pre-processing, Proposed method, Pre-training, Post-processing)

\subsection{T1: SiamRETFound for longitudinal change detection}

\begin{figure}[!hbt]
\begin{center}
\includegraphics[width=\linewidth]{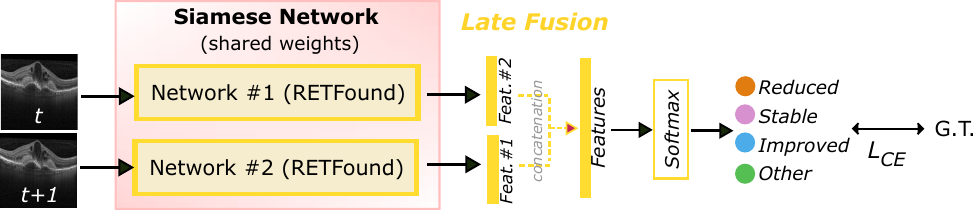}
\end{center}
\caption{SiamRETFound approach for longitudinal change detection in retinal OCT.}
\label{fig:method_scheme_t1}
\end{figure}

We propose SiamRETFound, a Siamese neural network with shared weights for evaluating change between retinal B-scans from two visits of the same patient (Fig.~\ref{fig:method_scheme_t1}). SiamRETFound uses a late fusion approach by extracting and concatenating two feature representations, one for each B-scan. The resulting feature vector, which integrates meaningful features from both time points, is then processed by a classification head to obtain the change detection. Ultimately, SiamRETFound learns to assess which differences between the extracted features are relevant to identify clinically meaningful changes.

%The backbone for feature extraction is the encoder part of the RETFound model, a foundation model pre-trained in retinal OCTs using a masked autoencoder strategy. 
The backbone for feature extraction consists of a pair encoders of the RETFound model, a foundation model pre-trained on retinal OCTs using a masked autoencoder strategy~\cite{Zhou2023}. 
The encoder is a large vision Transformer (ViT) with 24 Transformer blocks and embedding vector size of 1024.
%The RETFound model was pretrained on a total of 736442 OCT central B-scans from two datasets:
%\begin{inparaenum}[1)]
%\item MEH-MIDAS, composed of Topcon and Cirrus OCTs from 37401 diabetic patients from Moorfields Eye Hospital
%\item dataset from Kermany \textit{et al.}~\cite{Kermany2018}, with Heidelberg Spectralis OCTs from 4686 patients (Normal, iAMD, nAMD and DME classes).
%\end{inparaenum}
The classification head has a fully connected (FC) layer with 256 neurons and a ReLU activation, 25\% dropout, and a FC with 4 output neurons (number of classes). 

%Late fusion is applied, with the concatenation of the features from both images prior to the classification head. 

%The backbone chosen consists of the RETFound model, a foundation model trained for OCT classification tasks in OCT~\cite{Zhou2023}. 
%RETFound consists of a masked autoencoder, which consists of an encoder and a decoder. The encoder is a large vision Transformer (ViT) with 24 Transformer blocks and embedding vector size of 1024, and the decoder is a small ViT with 8 Transformer blocks and embedding vector size of 512. The objective during training was to reconstruct OCT images from highly masked versions of these images. %85\% mask ratio 
%The RETFound model was pretrained on a total of 736442 OCT central B-scans from two datasets:
%\begin{inparaenum}[1)]
%\item MEH-MIDAS, composed of Topcon and Cirrus OCTs from 37401 diabetic patients from Moorfields Eye Hospital
%\item dataset from Kermany \textit{et al.}~\cite{Kermany2018}, with Heidelberg Spectralis OCTs from 4686 patients (Normal, iAMD, nAMD and DME classes). \hl{I also mention kermany in Datasets so check repetition}
%\end{inparaenum}
%The 24 Transformer blocks, comprising multiheaded self-attention and multilayer perceptron, 
%(256$\times$256 pixels)

The full siamese network is first pretrained on a public OCT dataset using a simulated binary change detection task: \textit{change/no change}. 
The network is then finetuned on the MARIO training data for classification of the target classes.

\subsubsection{Datasets and Evaluation} \label{sec:data_task1}
For training and evaluating our model we used the MARIO challenge task 1 development data. This is composed of a training set for which ground truth labels are provided and a validation set for which labels are kept hidden and that is used for ranking participants. 
The training set 14496 has B-scan pairs from 68 patients, with up to 10 visits per patient. The validation set has 7010 pairs  from 34 patients.
All OCT volumes were acquired with Spectralis device and longitudinal 3D volumes were anatomically aligned with the follow-up mode from the device.

The public Kermany dataset~\cite{Kermany2018} used for pretraining consists of 100,000 images from patients belonging to one of the following classes: healthy, intermediate AMD, nAMD, and diabetic macular edema (DME). Surrogate binary labels \textit{change/no change} were inferred from the disease classes: if both images have the same disease, the pair was attributed to \textit{no change}, otherwise to \textit{change}.

%Evaluation metrics
The evaluation metrics reported in this work include the official MARIO challenge metrics: F1-score (micro-averaging), Specificity and Rk-correlation coefficient (i.e., Mathews correlation coeff. for multiclass setting), along with Balanced accuracy and Cohen's Kappa Coefficient ($k$) that we added.

\subsubsection{Training details}
The MARIO challenge training set was split patient-wise in 90\% for training and 10\% for validation. To generate the pretraining data from the Kermany dataset, random pairs are taken from the original dataset, for a total of 300,000 image pairs. 
Both pretraining and finetuning follow the same scheme. Training-time image augmentation is applied, ensuring the same augmentation on both images of a pair. Augmentations include resize-crop (minimum of 85\% image size), rotation (up to 15\textdegree) and horizontal flipping.  Images are resized to 224$\times$224 pixels and the intensities are normalized by the ImageNet mean and standard deviation. 
Batches are balanced class-wise during training to compensate for the class imbalance.
A warm-up is done in the beginning of the training with classification neurons frozen for 50 epochs.
%, then all neurons are unfreeze for the remaining of the training.
The Adam optimizer with an initial learning rate (LR) of 0.001 with an exponential decay was used, and the training loss was the cross-entropy loss. 
SiamRETFound was finetuned for up to 100 epochs with early stopping based on the average of the challenge evaluation metrics on the held out validation set. 

\subsubsection{Ensembling}

%For the MARIO challenge ranking specifically, we used an ensembling strategy. In particular, we combined the prediction probabilities of 10 different models:
For the MARIO challenge ranking specifically, we used an ensembling strategy by combining the prediction probabilities of 10 different models:
%Ensembling of models is known to often outperform single-model approaches, and thus we tested an ensembling of different models based on their soft predictions (i.e., output probabilities for each class). 
%We combined the predictions from 10 different models: 
\begin{inparaenum}[1)]
\item 5 SiamRETFound-based models (with different training settings: optimizers, LR, augmentation schemes);  
\item 5 Shifted WINdows (SWIN)\cite{liu2021swin} transformer-based models (5 folds). 
To obtain a prediction for a test image we took the mean of all models' predictions per class, and chose the class with the maximum-score.%, and compute the maximum in order to get the predicted class. 
\end{inparaenum}
% ------
% The Swin transformer-based approach consisted in supervised pretraining for disease classification on public datasets (OCTID~\cite{gholami2020octid}, OCTDL~\cite{kulyabin2024octdl}, and Kermany~\cite{Kermany2018}), then for biomarker detection on OLIVES data~\cite{prabhushankar2022olives}. The model is finally
% finetuned on MARIO data (5 folds) \hl{TODO check Marzieh}

In our SWIN-based models, we first trained a SWIN transformer sequentially on three publicly available OCT datasets: OCTID~\cite{gholami2020octid}, OCTDL~\cite{kulyabin2024octdl}, and Kermany~\cite{Kermany2018}. Second, we finetuned the model for multi-label biomarker detection using the OLIVES dataset~\cite{prabhushankar2022olives}
\footnote{Same splits as the IEEE SPS VIP Cup 2023: Ophthalmic Biomarker Detection.}. This method ensures that the model is exposed to a broad spectrum of diseases, enhancing its ability to detect critical biomarkers in B-scans, crucial for change detection.
%\footnote{We used the splits from the IEEE SPS VIP Cup 2023: Ophthalmic Biomarker Detection.}. This method ensures that the model is exposed to a broad spectrum of diseases, enhancing its ability to detect critical biomarkers in B-scans, crucial for change detection.

% % ====================================================================

\subsection{T2: WAsserstein distance for RetInal disease Ordinality modeling (WARIO)}
Our solution to T2 has 3 main components: masked-autoencoder based pretraining, finetuning with a novel approach to ordinal classes, and postprocessing to ensure volume level consistency based on individual B-scan level predictions.

\paragraph{Pretraining} Pretraining is an important step of the current deep learning pipeline. In AMD disease progression tasks, numerous works showed the benefit of pretraining~\cite{chakravarty2024morph,emre2022tinc}. Recently, masked auto-encoders~\cite{he2022masked} (MAE) achieved the state-of-the-art in vision tasks. Following this trend, we pretrain our models with MAE by masking 75\% of B-scans (Fig.~\ref{fig:mae}) from a combined dataset of T1 and T2.

\paragraph{Finetuning} 
Addressing T2 effectively requires accounting for (i) the class imbalance due to large number of \textit{Stable}, and (ii) the ordinal relationship between the classes (\textit{Reduced}-\textit{Stable}-\textit{Worsened}). While the class imbalance poses a challenge, the ordinality provides an opportunity for additional training signal. For the class imbalance, we adapted Focal Loss~\cite{lin2017focal}, which dynamically calculates weights for hard and easy examples, so that hard to classify samples will contribute more:% to the loss:%. In an imbalanced classification problem, it is defined as:
%Addressing T2 effectively should take into account two important aspects: (i) the class imbalance due to large number of \textit{Stable}, and (ii) the ordinal relationship between the classes (\textit{Reduced}-\textit{Stable}-\textit{Worsened}). While the class imbalance poses a challenge, the ordinality provides an opportunity for additional training signal. For the class imbalance, we adapted Focal Loss~\cite{lin2017focal}. It dynamically calculates weights for hard and easy examples, so that hard to classify samples will contribute more to the loss:%. In an imbalanced classification problem, it is defined as:
\begin{equation}
    \label{eq:foc}
   \ell_\text{focal} = -\alpha \sum_{i=0}^{C-1} y_i (1 - \hat{p}_i)^\gamma \log(\hat{p}_i),
\end{equation}
where $C$ is the number of classes, and $\gamma$ is the focusing hyper-parameter for hard to classify samples. Additionally we undersample the majority class (\textit{Stable}) during finetuning.

For the ordinality we use discrete earth-mover's distance~\cite{hou2016squared,talebi2018nima} (EMD, also known as Wasserstein-distance). Unlike the cross-entropy loss, it takes inter-class relations (e.g. ranking) into account when calculating the loss. In T2, the classes are ordered by their definition, as \textit{Reduced}-\textit{Stable}-\textit{Worsened}. EMD calculates cumulative mass need to be moved to make one distribution similar to another:

\begin{equation}
\label{eqn:emd}
\ell_\text{EMD} = \left( \frac{1}{C} \sum_{i=0}^{C-1} |\mbox{CDF}_y(i) - \mbox{CDF}_{\hat{p}}(i)|^2 \right)^{1/2},
\end{equation}
where $\mbox{CDF}_p()$ is the cumulative distribution function calculated from the class probabilities or one-hot encoded target vector. The final training loss is the combination of these two loss terms with equal weighting. 
\paragraph{Postprocessing (Ensembling)} 
After pretraining and finetuning on 3-fold splits, the predictions are ensembled as follows. We make a single prediction from different training split folds in a B-scan level, and then an eye level prediction from available B-scans of a particular OCT volume. In general, the trained networks tend to predict \textit{Stable} due to the imbalance. In order to alleviate this, the final prediction from the 3 outputs is set to class \textit{Stable} if only all 3 models predict \textit{Stable}, otherwise it is set to the majority vote.
%After pretraining and finetuning on 3-fold splits, a post-processing step is needed to make a final prediction in an ensemble manner. The pipeline consists of 2 main parts: making a single prediction from different training split folds in a B-scan level, and making an eye level prediction from available B-scans of a particular OCT volume. In general, the trained networks tend to predict \textit{Stable }class due to the imbalance. In order to alleviate this, the final prediction from the 3 outputs is set to class \textit{Stable} if only all 3 models predict \textit{Stable}, otherwise it is set to the majority vote.

In T2, the labels are provided at the OCT volume level, i.e. B-scans from a particular visit scan of a patient have the same label. To ensure this consistency in our predictions, we apply a majority voting. In order to avoid only predicting the majority \textit{Stable} class, a volume level label is set to \textit{Stable} if at least 80\% (the class ratio) of the B-scan level predictions are \textit{Stable}. If not, volume level label is the majority prediction. Then we set all B-scans of that volume to a single label, and reported the predictions in this manner.

\subsubsection{Datasets and Evaluation} 
T2 training data consisted of 8082 B-scans from 330 volumes for 61 patients. The provided validation split contained 3822 B-scans from 163 volumes for 29 patients. All OCT volumes are Spectralis and consecutive 3D volumes are registered. For metrics, the official MARIO challenge metrics: F1-score (micro-averaging), Specificity, Quadratic-weighted Kappa and Rk-correlation coefficient, and additional Balanced accuracy are used.

\subsubsection{Training details}
We use ViT~\cite{dosovitskiy2021an} base model with a single FC prediction head. It has 12 MHSA blocks with 12 parallel heads, embedding size of 768 and $16 \times 16$ patch size. The prediction head uses global average pooled embedding features instead of the CLS token. We first run MAE pretraining for 800 epochs with a decayed LR of 3E-4, weight decay of 1E-2, and batch size of 256. We use AdamW as the optimizer with its Beta2 hyperparameter set to 0.95. We then finetune end-to-end for 200 epochs with a warm-up of 20 epochs, a LR of 2.5E-4 and a weight decay of 1E-4. We used AdamW as the optimizer with Beta2 of 0.999. We only use 2D B-scans and considered each B-scan i.i.d (even from the same patient) during training. Training augmentations include translation, small rotation, and horizontal flipping. 
Each image is resized to $224 \times 224$ pixels.
%We use ViT~\cite{dosovitskiy2021an} base model with a single FC prediction head. It has 12 MHSA blocks with 12 parallel heads and embedding size of 768. The patch size is set to $16 \times 16$. The prediction head uses global average pooled embedding features instead of CLS token. We first run MAE pretraining for 800 epochs with a decayed learning rate of 3E-4, weight decay of 1E-2, and batch size of 256. We use AdamW as the optimizer with its Beta2 hyperparameter set to 0.95. Then for the finetuning, we follow an end-to-end fashion for 200 epochs with a warm-up of 20 epochs, a learning rate of 2.5E-4 and a weight decay of 1E-4. We used AdamW as the optimizer with Beta2 of 0.999. We only use 2D B-scans and considere each B-scan i.i.d even the ones from same patient during training. For training augmentations, translation, small rotation, and horizontal flipping are used. Each image is resized to $224 \times 224$ pixels.

%
% ====================================================
% ====================================================

\section{Results and Discussion}

%Table~\ref{tab:metrics} shows the performance of the proposed models for the task of longitudinal change detection (T1) and severity change prediction (T2) on the MARIO challenge data, for the different evaluation metrics.  
%Confusion matrices are shown in Appendix~\ref{app:conf_mat}. 
%\TEc{This intro could be deleted completely}

\subsection{T1: Longitudinal change detection }

\begin{figure}[!htb]
\centering
\begin{minipage}{.44\linewidth}
  \centering
  \centerline{\includegraphics[width=\linewidth]{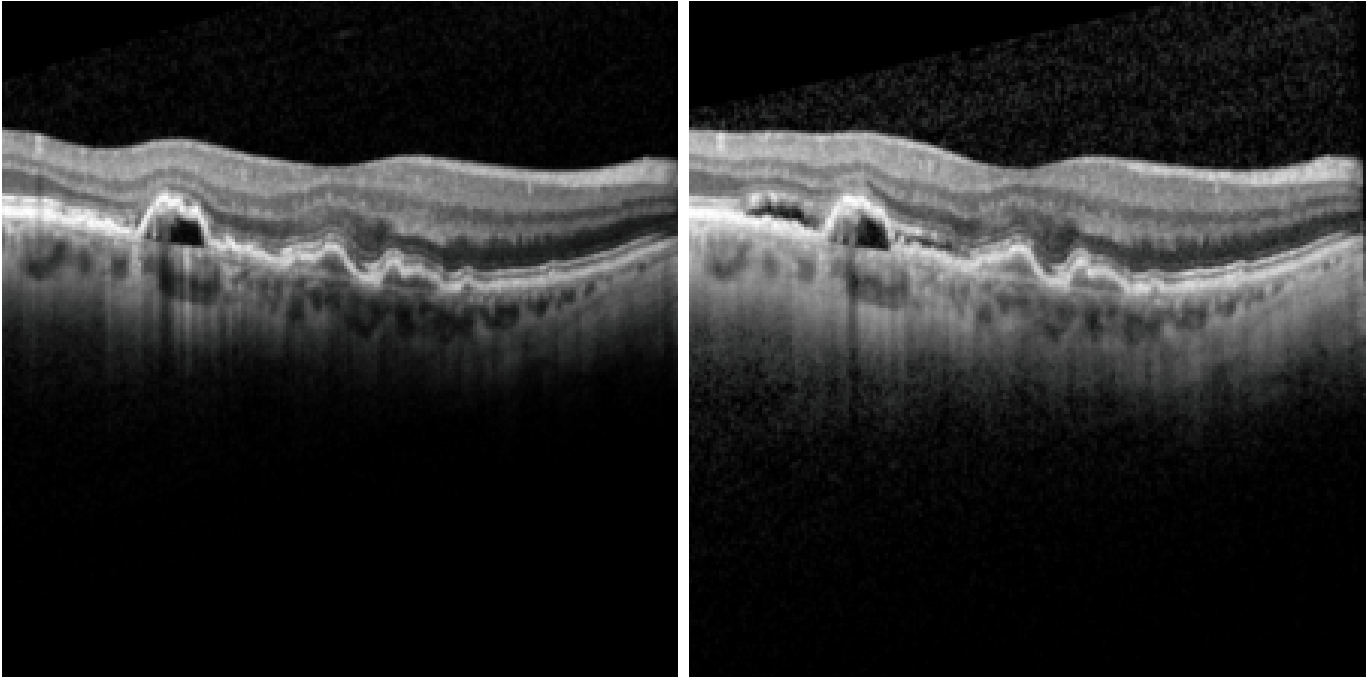}}
  \centerline{(a) GT: \textit{Worsened}, P: \textit{Worsened}. }%\medskip
\end{minipage}
\begin{minipage}{.44\linewidth}
  \centering
  \centerline{\includegraphics[width=\linewidth]{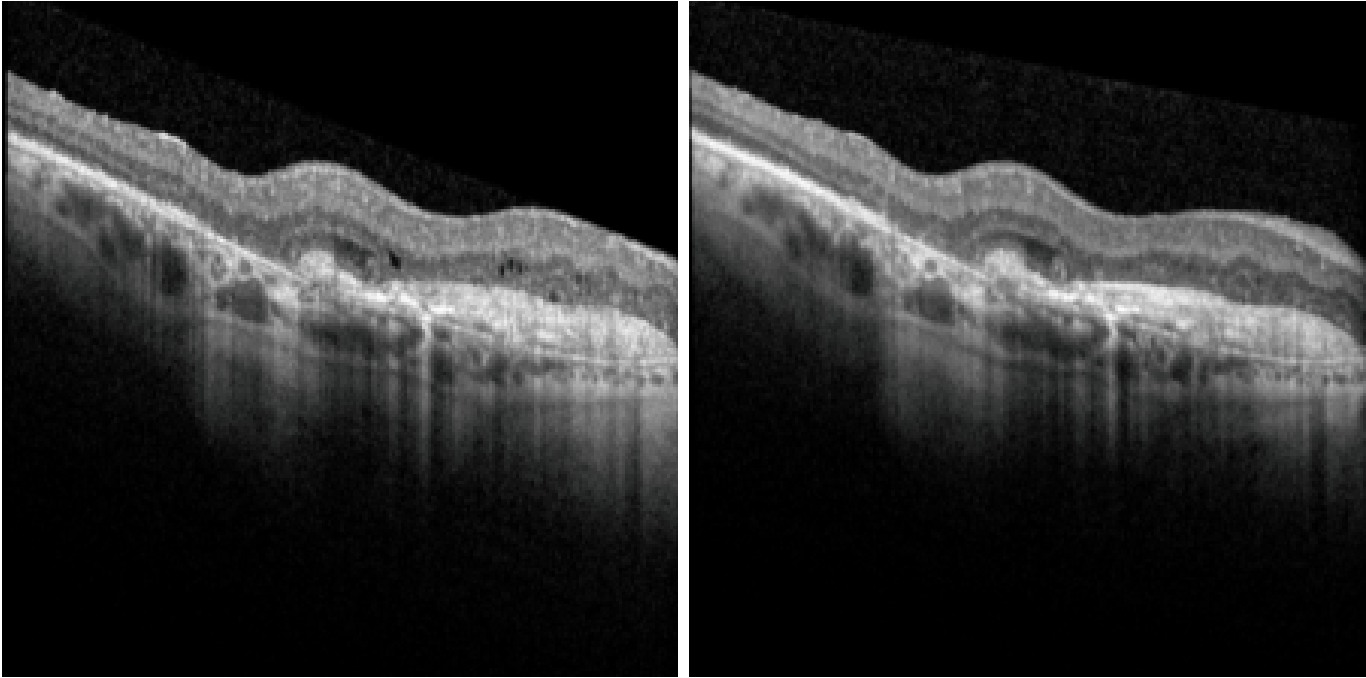}}
  \centerline{(b) GT: \textit{Reduced}, P: \textit{Reduced}.}%\medskip
\end{minipage}
%\begin{minipage}{.43\linewidth}
%  \centering
%  \centerline{\includegraphics[width=\linewidth]{figures/exs_preds_T1/im4_gt_worsened_pred_worsened_cut.png}}
%  \centerline{(c) GT: \textit{Worsened}, P: \textit{Worsened}.}\medskip
%\end{minipage}
\begin{minipage}{.44\linewidth}
  \centering
  \centerline{\includegraphics[width=\linewidth]{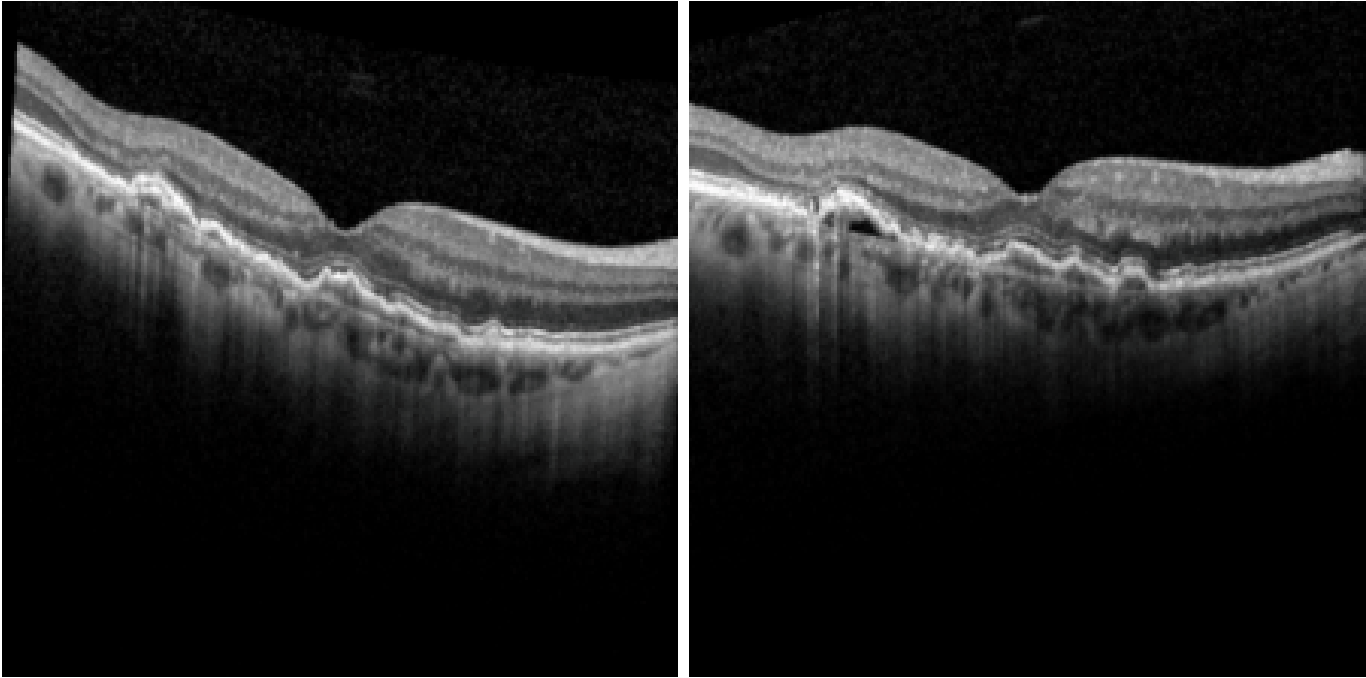}}
  \centerline{(d)  GT: \textit{Stable}, P: \textit{Worsened}.}%\medskip
\end{minipage}
\begin{minipage}{.44\linewidth}
  \centering
  \centerline{\includegraphics[width=\linewidth]{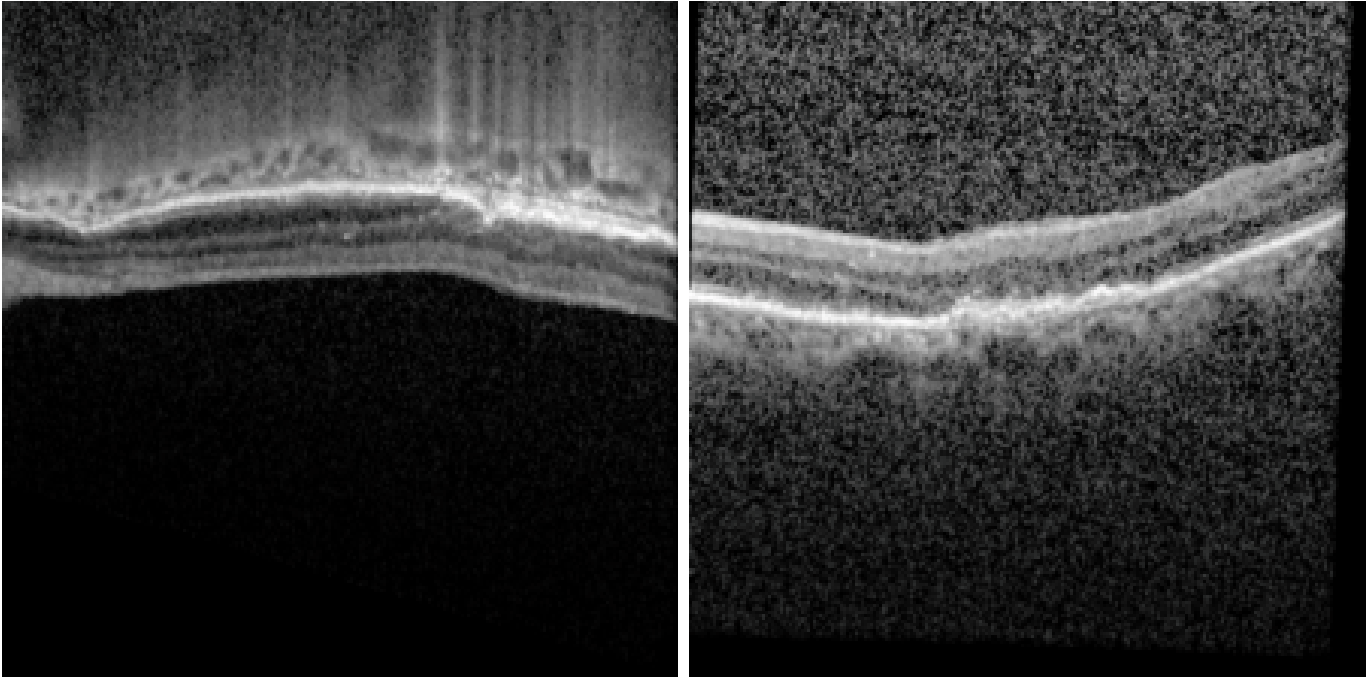}}
  \centerline{(e) GT: \textit{Other}, P: \textit{Other}.}%\medskip
\end{minipage}
\begin{minipage}{.44\linewidth}
  \centering
  \centerline{\includegraphics[width=\linewidth]{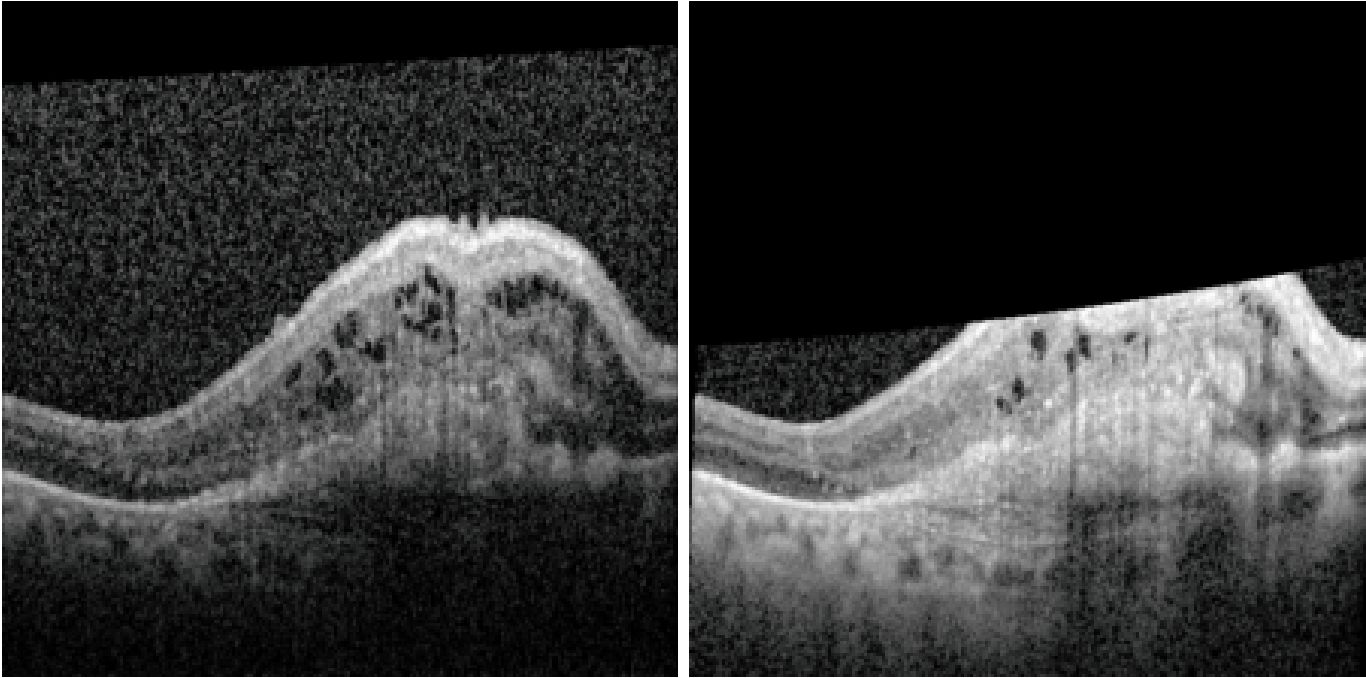}}
  \centerline{(f) GT: \textit{Other}, P: \textit{Other}.}%\medskip
\end{minipage}
\begin{minipage}{.44\linewidth}
  \centering
  \centerline{\includegraphics[width=\linewidth]{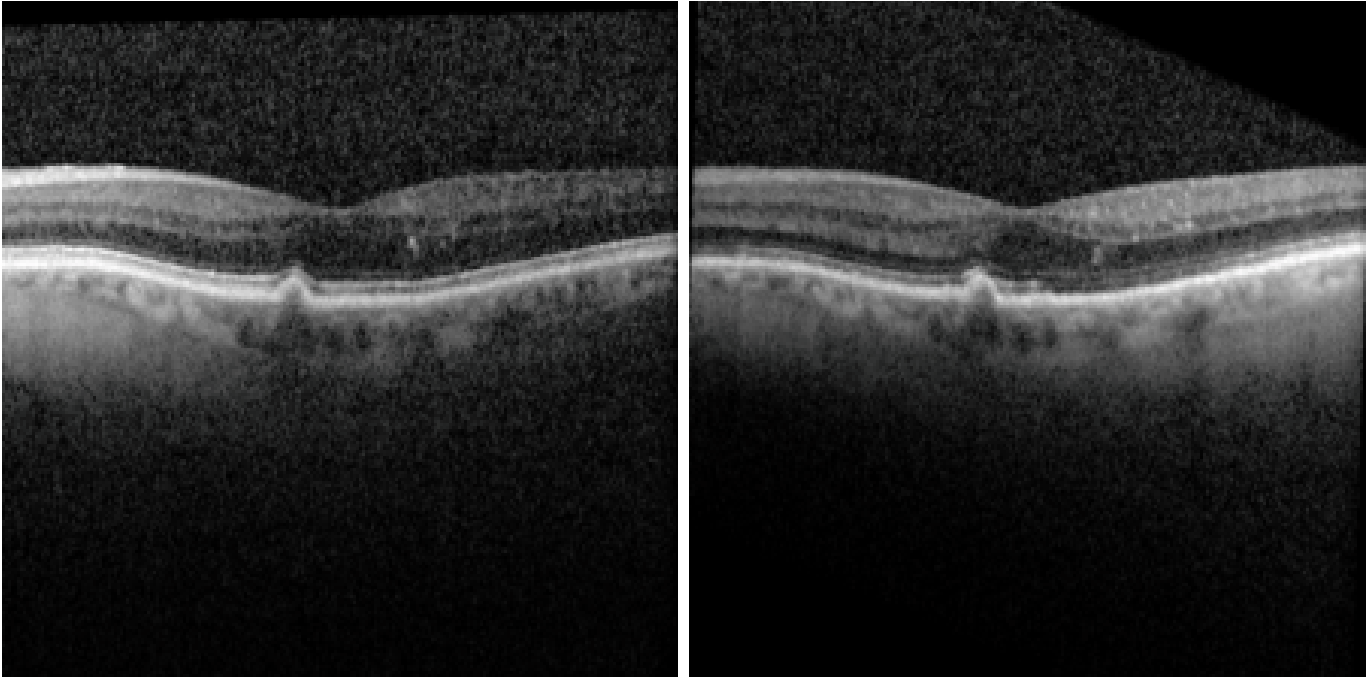}}
  \centerline{(g) GT: \textit{Other}, P: \textit{Stable}.}%\medskip
\end{minipage}
%\begin{minipage}{.46\linewidth}
%  \centering
%  \centerline{\includegraphics[width=\linewidth]{figures/exs_preds_T1/im1107_label_other_pred_improved_cut.png}}
%  \centerline{(h) GT: \textit{Other}, P: \textit{Reduced}.}%\medskip
%\end{minipage}
%\vspace{-0.5em}
\caption{Examples of SiamRETFound predictions (GT: ground truth, P: prediction).}
    \label{fig:exs_preds}
\end{figure}

%Examples of images and the corresponding predictions and ground truth are shown in Fig.~\ref{fig:exs_preds}. 
%Occlusion sensitivity maps are shown in Appendix~\ref{app:occ_sens}. 
% Discussion TASK 1 ------------------------------------------------------
We showed that SiamRETFound is able to correctly classify longitudinal change for the majority of the OCT pairs (Table~\ref{tab:metrics}). The model is able to capture very subtle changes (Fig.~\ref{fig:exs_preds}: (a),(b)). Specifically for the change classes (\textit{Reduced}, \textit{Stable} and \textit{Worsened}), which can be interpreted as an ordinal classification problem, we verified most errors occur between neighboring classes and little \textit{Reduced}/\textit{Worsened} cases are confounded (Appendix - Fig.~\ref{fig:conf_mat_t1}). For the class Other, the majority of the confusion occurs with \textit{Stable}, which can be attributed to the not so clear definition of \textit{Other} (e.g., how much noise or obscured area constitutes an uninterpretable pair) (Fig.~\ref{fig:exs_preds}: (g)). 

The occlusion sensitivity maps (Appendix - Fig.~\ref{fig:occ_maps}) suggest that the SiamRETFound model is capturing relevant features to classify change between image pairs, mostly focusing on fluid regions. 

Despite SiamRETFound approach being better in terms of overall classification  performance (inferred by the confusion matrix and the balanced accuracy and $k$ metrics), concerning the MARIO challenge metrics the ensemble version outperformed the single model prediction. 
On the challenge leaderboard in the development phase we ranked \ordinalnum{4} considering the mean of the evaluation metrics, with a score of 0.801 (scores of the top 3 teams: 0.828/0.805/0.804).

\begin{table}[!t]
\centering
\caption{Results of our methods in the MARIO challenge validation sets for T1 and T2. For T1, Kappa is Cohen's Kappa; for T2, it is Quadratic-weighted Kappa.}
\label{tab:metrics}
\resizebox{\textwidth}{!}{
\begin{tabular}{lccccc}
\toprule
\multirow{2}{*}{\textbf{Metric}} & \multicolumn{2}{c}{\textbf{Task 1 (T1)}} & \multicolumn{2}{c}{\textbf{Task 2 (T2)}} \\
\cmidrule(lr){2-3} \cmidrule(lr){4-5}
& \multicolumn{1}{p{2.5cm}}{\centering \textbf{SiamRETFound}} & \multicolumn{1}{p{2.5cm}}{\centering \textbf{Ensemble}} & \multicolumn{1}{p{2.5cm}}{\centering \textbf{WARIO}} & \multicolumn{1}{p{2.5cm}}{\centering \textbf{Ensemble}} \\
\midrule
\textbf{Balanced Acc.} & 0.713 & 0.691 & 0.336 & 0.485 \\
\textbf{Kappa$^{T2}$} & 0.642 & 0.656 & 0.133 & 0.223 \\
\textbf{F1 Score$^{T1,T2}$} & 0.817 & 0.833 & 0.720 & 0.705 \\
\textbf{Rk-Correlation$^{T1,T2}$} & 0.642 & 0.657 & 0.004 & 0.206 \\
\textbf{Specificity$^{T1,T2}$} & 0.917 & 0.911 & 0.666 & 0.722 \\
% Averageın üstüne çizgi çek, hrule
\textbf{Average$^{(*)}$} & 0.792 & 0.801 & 0.351 & 0.469 \\
\bottomrule
\end{tabular}
}
\end{table}

% ====================================================

\subsection{T2: Prediction of AMD evolution within 3 months}

% Discussion TASK 2 ------------------------------------------------------
In Task 2, the goal is to detect the change in AMD severity in 3 months from a provided past visit scan. We found that, it is crucial to have a strong pretraining step and correct loss terms. Only the combination of focal loss and EMD loss (WARIO) prevent the network from only predicting the \textit{Stable} class. In Table~\ref{tab:metrics}, it is clear that WARIO is still heavily biased towards the majority class, highlighting the importance of a postprocessing step. Clinically we know that AMD change is a OCT volume level assessment from which we concluded that the B-scan level predictions need to be consistent along an OCT volume. We enforced this consistency in the postprocessing step combined with an ensemble of 3-fold data split. Even though F1-score dropped slightly, WARIO improved along other metrics. It is important to highlight that postprocessing improved Rk-correlation the most which is generally used in imbalanced classification problems. At the end, WARIO ranked \ordinalnum{2} in the challenge learderboard.

% ====================================================
\section{Conclusions}

MICCAI'24 MARIO challenge aims to model AMD severity change in 2 setups: (i) predicting the change by comparing B-scans from two time points, (ii) predicting the change after 3 months from a single B-scan.
We proposed SiamRETFound, a siamese-based approach that was able to effectively classify longitudinal change in OCT pairs, even for very subtle cases. The proposed model captured relevant features to classify change between B-scan pairs, mostly focusing on exudative regions. 
Predicting the change in OCTs from a single past visit is extremely challenging task. Our method, WARIO, uses focal loss for class imbalance and EMD loss to exploit ordinal relation in the severity change classes, preceded by a strong MAE based pretraining. The proposed methods show potential to facilitate the clinical workflow on nAMD diagnosis from retinal OCTs and allow timely and thus more effective treatment planning.

\bibliographystyle{splncs04}

\bibliography{references.bib,bibtaha,refs_swin}

%\hl{TODO reduce number of authors (eg Retfound)}

% ===========================================
% ===========================================
% From here appendix
\clearpage
\appendix
\section{Appendix}

% ===========================================
%\subsection{Confusion matrices}
%\label{app:conf_mat}
\begin{figure}[!htb]]
\centering
\begin{minipage}{.40\linewidth}
  \centering
  \centerline{\includegraphics[width=\linewidth]{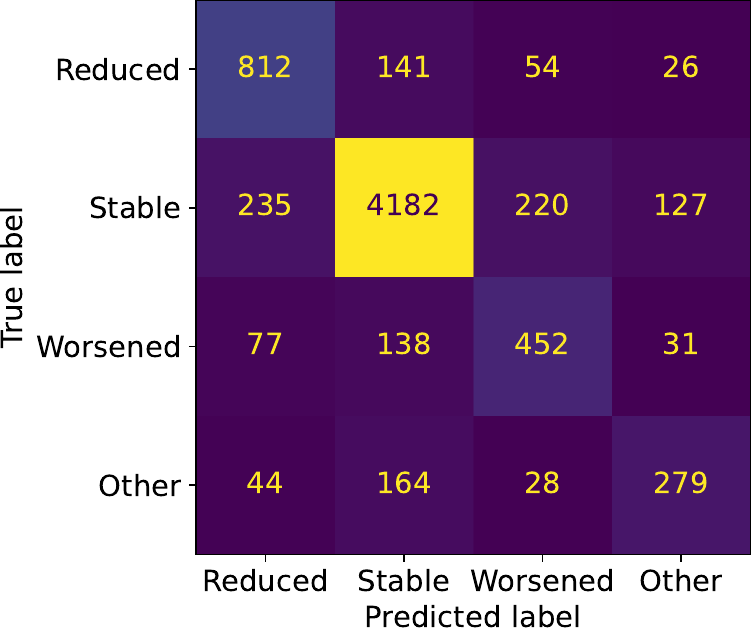}}
  \centerline{(a) Single model.}%\medskip
\end{minipage}
\begin{minipage}{.40\linewidth}
  \centering
  \centerline{\includegraphics[width=\linewidth]{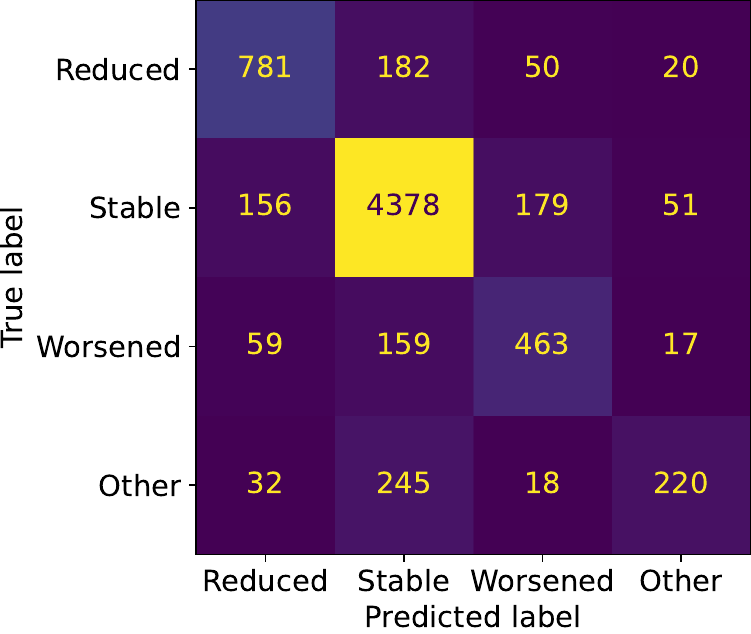}}
  \centerline{(b) Ensemble.}%\medskip
\end{minipage}
%\vspace{-0.5em}
\caption{Confusion matrices for Longitudinal change detection (MARIO Task 1).}
    \label{fig:conf_mat_t1}
\end{figure}

\vskip-2.5em
\begin{figure}[!htb]]
\centering
\begin{minipage}{.40\linewidth}
  \centering
  \centerline{\includegraphics[width=\linewidth]{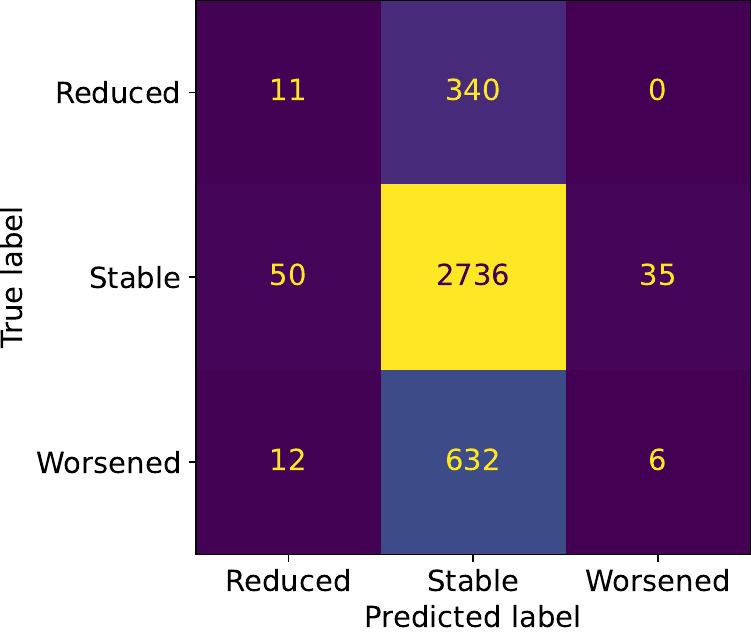}}
  \centerline{(a) Single model}%\medskip
\end{minipage}
\begin{minipage}{.40\linewidth}
  \centering
  \centerline{\includegraphics[width=\linewidth]{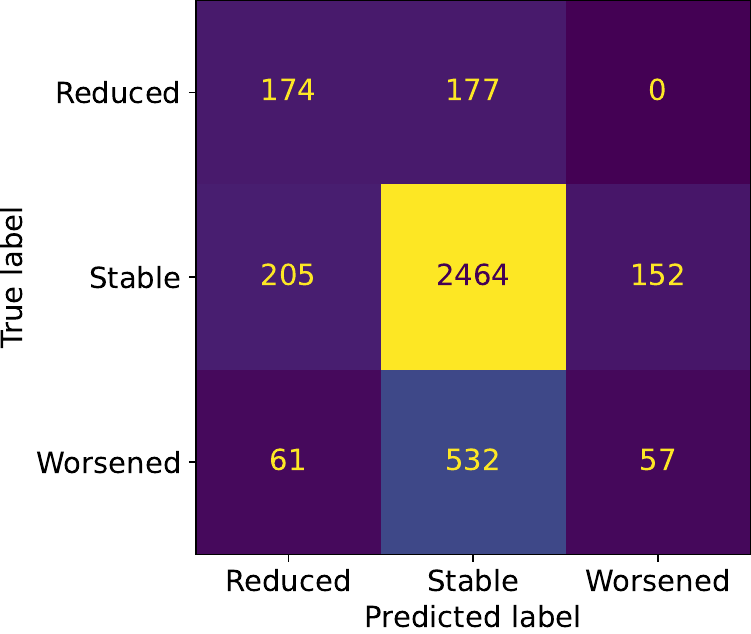}}
  \centerline{(b) Ensemble} 
\end{minipage}
%\vspace{-0.5em}
\caption{Confusion matrices for AMD evolution prediction (MARIO Task 2)} 
    \label{fig:conf_wario}
\end{figure}

% ===========================================
%\subsection{Occlusion sensitivity}
%\label{app:occ_sens}

%Occlusion sensitivity approach was applied for SiamRETFound in order to visualize the areas of the image pair that were more relevant for the networks' classification, i.e., the regions that were more important to determine the change between the two images within a pair. 

%To do this, patches of N x N pixels (N = 4) are occluded in both images of a given pair. The occluded images are passed through the Siamese NN and the softmax output is computed. The predicted probability for each occluded image for the originally predicted class (i.e., network prediction for the non-occluded image) is assigned to the occlusion map at the area of the occluded patch. 
%The intuition behind this approach is that the output probability decreases when a patch relevant for the classification is occluded. 
%\hl{justify patch size?}
%From Li2020: The patch sizes were selected empirically based on the tradeoff between spatial sensitivity and noise.

\begin{figure}[!htb]
\centering
\begin{minipage}{\linewidth}
  \centering
  \centerline{\includegraphics[width=.85\linewidth]{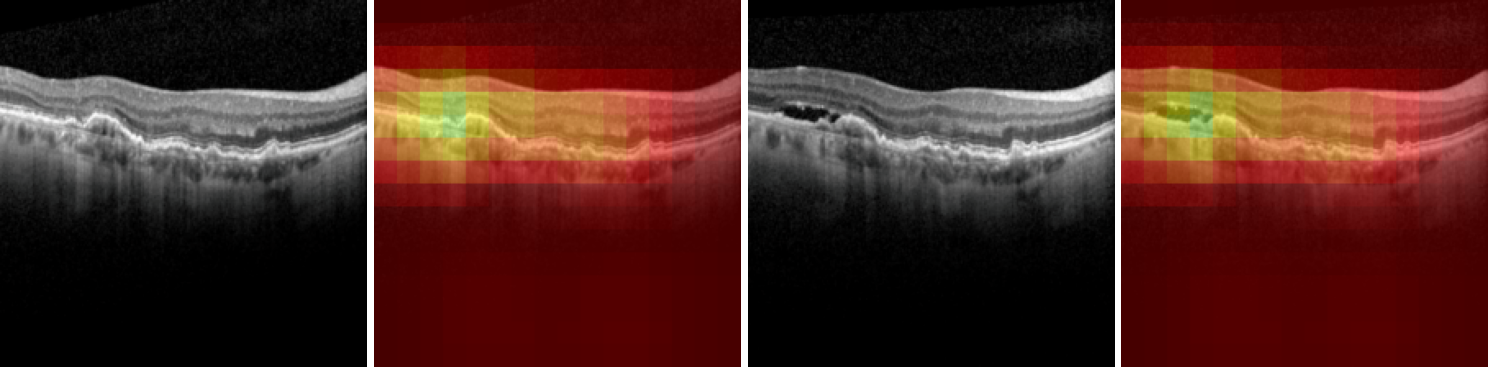}}
  \centerline{(a) GT: \textit{Worsened}, P: \textit{Worsened}. }%\medskip
\end{minipage}
\begin{minipage}{\linewidth}
  \centering
  \centerline{\includegraphics[width=.85\linewidth]{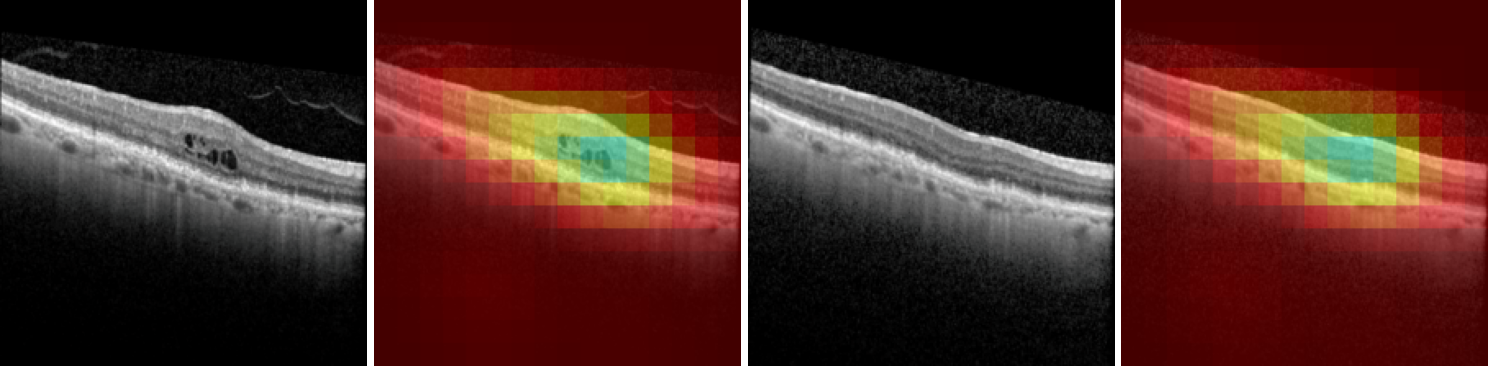}}
  \centerline{(b) GT: \textit{Reduced},, P: \textit{Reduced}.}%\medskip
\end{minipage}
\begin{minipage}{\linewidth}
  \centering
  \centerline{\includegraphics[width=.85\linewidth]{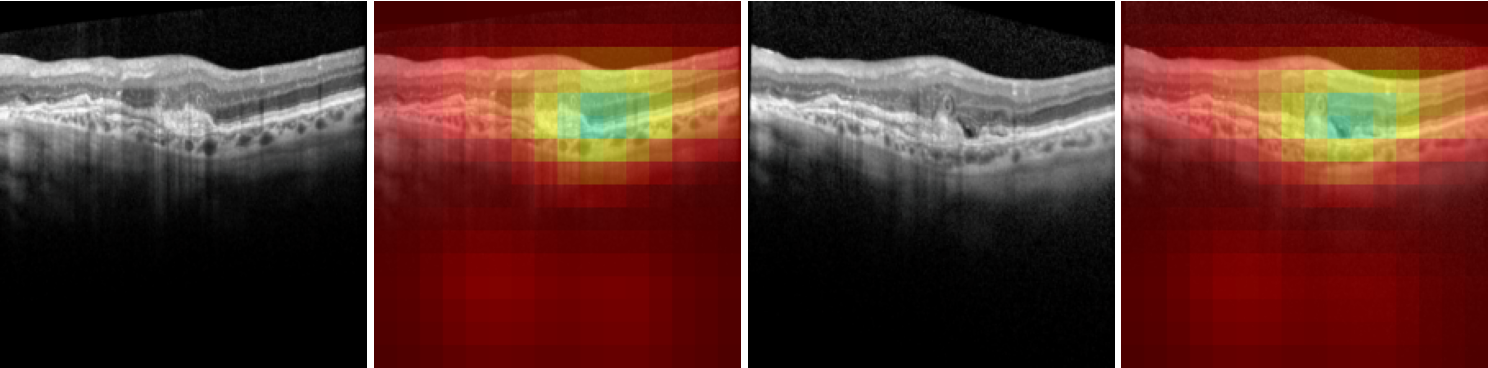}}
  \centerline{(c) GT: \textit{Worsened}, P: \textit{Worsened}.}%\medskip
\end{minipage}
\begin{minipage}{\linewidth}
  \centering
  \centerline{\includegraphics[width=.85\linewidth]{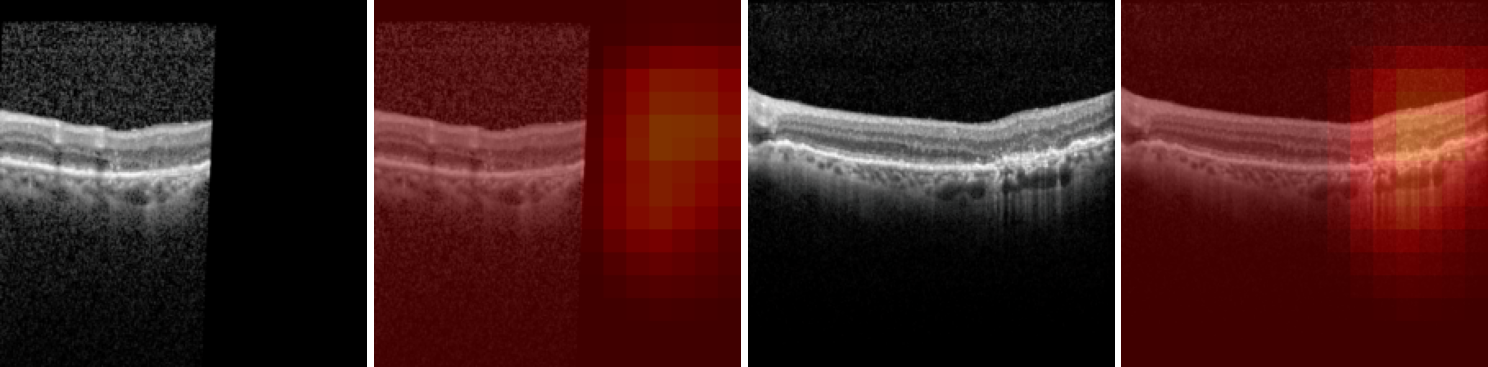}}
  \centerline{(d) GT: \textit{Other}, P: \textit{Other}.}
\end{minipage}
\caption{Occlusion map sensitivity. %Examples (a)-(c) show the model is focusing more on regions relevant for diagnosis; case (d) shows an example where the most relevant region for classification was not the expected (not focusing the fluid region). 
Map values are from 0 (blue) to 1 (red). Lower values indicate the occluded region impacted more the final prediction.}
    \label{fig:occ_maps}
\end{figure}

% ===========================================

%\subsection{MAE Result}
\begin{figure}[hbt]
\begin{center}
\includegraphics[width=0.85\linewidth]{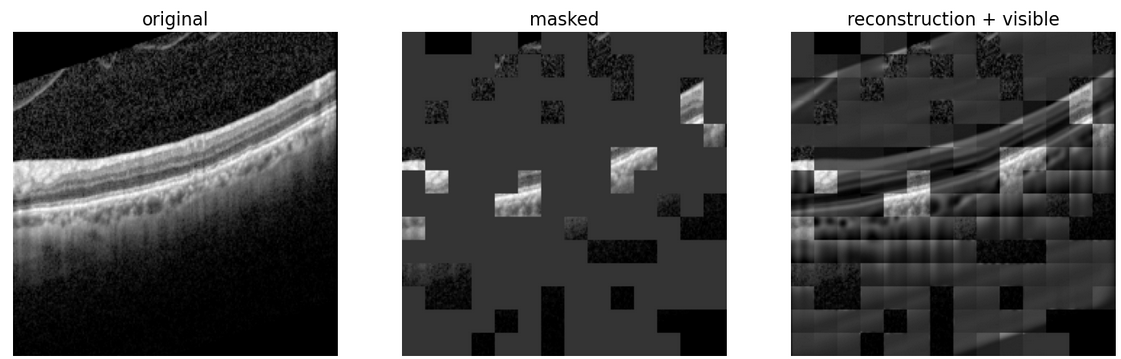}
\end{center}
\caption{MAE pretraining with masking and reconstruction.}
\label{fig:mae}
\end{figure}

\end{document}

%% file: mathcommends.tex
\usepackage{amsmath,amsfonts,bm}

\def\eqref#1{equation~\ref{#1}}
\def\1{\bm{1}}

\DeclareMathAlphabet{\mathsfit}{\encodingdefault}{\sfdefault}{m}{sl}
\SetMathAlphabet{\mathsfit}{bold}{\encodingdefault}{\sfdefault}{bx}{n}